# Gibbs Paradox and the Concepts of Information, Symmetry, Similarity and Their Relationship

**Shu-Kun Lin**

Molecular Diversity Preservation International (MDPI), Matthaeusstrasse 11, CH-4057 Basel, Switzerland; Tel. (+41) 79 322 3379; Fax: (+41) 61 302 8918; E-mail: lin@mdpi.org; http://www.mdpi.org/lin

(A commentary for Entropy journal)

We are publishing volume 10 of *Entropy*. When I was a chemistry student I was facinated by thermodynamic problems, particularly the Gibbs paradox. It has now been more than 10 years since I actively published on this topic [1-4]. During this decade, the globalized *Information Society* has been developing very quickly based on the Internet and the term "information" is widely used, but what is information? What is its relationship with entropy and other concepts like symmetry, distinguishability and stability? What is the situation of entropy research in general? As the Editor-in-Chief of *Entropy*, I feel it is time to offer some comments, present my own opinions in this matter and point out a major flaw in related studies.

**Definition of Information**

We are interested in the definition of information in the context of information theory. It is a surprise that a clear definition of the concept of "information" cannot be found in information theory textbooks. "Entropy as a measure of information" is confusing. I would like to propose a simple definition of information:

> Information ($I$) is the amount of the data after data compression.

If the total amount of data is $L$, entropy ($S$) in information theory is defined as information loss, $L = S + I$. Let us consider a 100GB hard disk as an example: $L = 100\text{GB}$. A formatted hard disk will have $S = 100\text{GB}$ and $I = 0$. Similar examples for defining information as the amount of data after compression are given in [5]. Based on this definition of information and the definition that (information theory) entropy is expressed as information loss, $S = L - I$, or in certain cases when the absolute values are unknown, $\Delta S = \Delta L - \Delta I$, I was able to propose three laws of information theory [5]:

> *The first law of information theory:* the total amount of data $L$ (the sum of entropy and information, $L = S + I$) of an isolated system remains unchanged.

*The second law of information theory:* Information (*I*) of an isolated system decreases to a minimum at equilibrium.

*The third law of information theory:* For a solid structure of perfect symmetry (e.g., a perfect crystal), the information *I* is zero and the (information theory) entropy (called by me as static entropy for solid state) *S* is at the maximum.

If entropy change is information loss, $\Delta S = -\Delta I$, the conservation of *L* can be very easily satisfied, $\Delta L = \Delta S + \Delta I = 0$. Another form of the second law of information theory is: the entropy *S* of the universe tends toward a maximum. The second law given here can be taken as a more general expression of the Curie-Rosen symmetry principle [5,6]. The third law given here in the context of information theory is a reflection of the fact that symmetric solid structures are the most stable ones.

Another reason to discuss structural stability and process spontaneity in the field of information theory is that, to the dismay of many thermodynamicists or physical chemists, not all the chemical processes are related to energy minimization and some processes may have nothing to do with energy; examples are some of the complicated processes of molecular recognition or very simple mixing processes where the process spontaneity can be elegantly and pertinently considered only in information theory. By (information theory) entropy (also sometimes known as information theory entropy, informational entropy, information entropy, or Shannon entropy), we mean it is a dimensionless logarithmic function in information theory. The so-called static entropy [5] as an (information theory) entropy is a nonthermodynamic entropy. Unlike thermodynamic entropy, which is also a special kind of (information theory) entropy, other kinds of entropy are not functions of temperature *T* and they are not necessarily related to energy.

Let us consider mixing processes to clarify the difference between thermodynamic entropy and the other kinds of (information theory) entropy.

**Gibbs Paradox**

Ten year ago, Edwin Thompson Jaynes (5 July 1922 – 30 April 1998), a physicist well-known for his contributions to information theory, was exchanging e-mails with me about the Gibbs Paradox. In his paper [7] he had complained that Gibb's 1902 book *Statistical Mechanics* "is the work of an old man in rapidly failing health, with only one more year to live. Inevitably, some arguments are left imperfect and incomplete towards the end of the work". Gibbs provided a famous explanation of the paradox bearing his name at the end of this book. After reading my paper [1] on Gibbs paradox, he was asked by me to provide his most recent opinion about the Gibbs paradox, and Jaynes only told me that he was under intensive nursery care (possibly in a hospital) as his immediate answer. His situation was that of 1902 of Gibbs. About two months or so later, Jaynes's colleague checked his e-mail inbox and told me in an e-mail that he had passed away.

Since then, several papers and a special issue on this topic have been published in *Entropy*. The following paragraphs are basically what I would like to contribute to the Wikipedia article *Gibbs Paradox*.

Today, the debate around this problem continues [8-10]. When Gibbs' paradox is discussed, the correlation of the entropy of mixing with similarity is always very controversial and there are three very different opinions regarding the entropy value as related to the similarity (Figures 1a, 1b and 1c).

Similarity may change continuously: similarity $Z = 0$ if the components are distinguishable; similarity $Z = 1$ if the parts are indistinguishable.

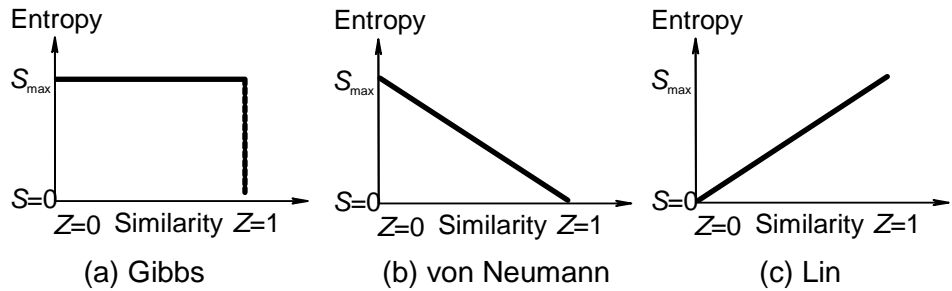

**Figure 1.** Entropy of mixing and similarity.

Entropy of mixing does not change continuously in the Gibbs paradox (Figure 1a). Let us recall that entropy of mixing of liquids, solids and solutions has been calculated in a similar fashion and the Gibbs paradox can be applied to liquids, solids and solutions in condensed phases as well as the gaseous phase.

In his book *Mathematical Foundations of Quantum Mechanics* [11] John von Neumann provided, for the first time, a resolution to the Gibbs paradox by removing the discontinuity of the entropy of mixing: it decreases continuously with the increase in the property similarity of the individual components (See Figure 1b). On page 370 of the English version of this book [11] it reads that " ... this clarifies an old paradox of the classical form of thermodynamics, namely the uncomfortable discontinuity in the operation with semi-permeable walls... We now have a continuous transition."

Another entropy continuity relation has been proposed by me [1-4], based on information theory considerations, as shown in Figure 1c. A calorimeter might be employed to determine the entropy of mixing and to either verify the proposition of Gibbs paradox or to resolve the Gibbs paradox. Unfortunately it is well-known that none of the typical mixing processes, whether they are in gaseous phase or in liquid or solid phases, have a detectable amount of heat and work transferred, even though a large amount of heat, up to the value calculated as $T\Delta S_T$ (where $T$ is temperature and $S_T$ is thermodynamic entropy), should have been measured and a large amount of work up to the amount calculated as $\Delta G$ (where $G$ is the Gibbs free energy) should have been observed. We may have to, rather reluctantly, accept the simple fact that the *thermodynamic entropy change of mixing of ideal gases is always zero*, whether the gases are different or identical (This conclusion might be taken as an experimental resolution of Gibbs paradox for ideal gases). This suggests that entropy of mixing has nothing to do with energy (heat $T\Delta S_T$ or work $\Delta G$). A mixing process may be a process of *information loss* which can be pertinently discussed only in the realm of information theory and entropy of mixing is an (information theory) entropy. Instead of calorimeters, chemical sensors or biosensors can be used to assess the information loss during the mixing process. Mixing 1 mol of gas A and 1 mol of a different gas B will have the increase of at most 2 bits of (information theory) entropy if the two parts of the gas container are used to record 2 bits of information ($I$). For mixing ideal gases, if the entropy is regarded as an (information theory) entropy, von Neumann's relation given in Figure 1b is valid.

For condensed phases, however, instead of the word "mixing", the word "merging" can be used for the process of combining several parts of substance originally in several containers. Then, it is always a merging process, whether the substances are very different or very similar or even the same. The conventional way of entropy of mixing calculation would predict that the mixing (or merging) process of different (distinguishable) substances is more spontaneous than the merging process of the same (indistinguishable) substances. However, this contradicts all the observed facts in the physical world where the merging process of the same (indistinguishable) substances is the most spontaneous one; immediate examples are spontaneous merging of oil droplets in water and spontaneous crystallization where the indistinguishable unite lattice cells ensemble together. More similar substances are more spontaneously miscible. The two liquids methanol and ethanol are miscible because they are very similar. Without exception, all the experimental observations support the entropy-similarity relation given in Figure 1c. It follows that the entropy–similarity relation of Gibbs paradox given in Figure 1a is questionable. A significant conclusion is that, at least in the solid state, the entropy of mixing is a negative value for distinguishable solids: mixing different substances will decrease the (information theory) entropy, and the merging of the indistinguishable molecules (from a large number of containers) to form a phase of pure substance has a great increase in static entropy—an (information theory) entropy [1-4]. Starting from a binary solid mixture, the process of merging 1 mol of molecules A to become one phase and merging of 1 mol of molecules B to form another phase leads to an (information theory) entropy increase of $2 \times 6.022 \cdot 10^{23} \text{bit} = 12.044 \cdot 10^{23} \text{bit} = 1.506 \cdot 10^{23} \text{Byte}$, where $6.022 \cdot 10^{23}$ is Avogadro's number; and there will be at most only 2 bits of information ($I$) left.

In conclusion, spontaneously mixed substances at gaseous state can be spontaneously separated at condensed phases (solid or liquid states), driving only by information loss or by the increase in (information theory) entropy. However, none of these typical pure mixing or separation processes are driving by free energy minimization and the free energy (or total amount of chemical potential) has no change during the processes of ideal mixture formation or ideal mixture separation. The thermodynamic entropy change for the formation of ideal mixtures of gases, liquids or solids is always zero.

There are several outstanding problems (such as symmetry breaking problem and related problems [8-10, 12-19]) which might be considered in information theory. I have introduced information theory concepts to the studies of structural stability and process spontaneity and tried hard to attack these problems myself before launching this journal 10 years ago and tried to present unambiguously my own main ideas. Now I am ready to continue to edit and publish many papers from other scientists in the following volumes. Contributions of papers are welcomed.

*Acknowledgements:* I am grateful to my long time colleague Dr. Derek McPhee for his collaboration and assistance. Derek corrected the English for this commentary.